\documentclass[modern]{aastex63}

\usepackage{graphicx} % Required for including pictures
\usepackage{amsmath}
\usepackage{ amssymb }

\usepackage{color}

\usepackage{float} % Allows putting an [H] in \begin{figure} to specify the exact location of the figure
\usepackage{wrapfig} % Allows in-line images such as the example fish picture

\usepackage[makeroom]{cancel} %allows strikethroughs
\usepackage{comment}

\usepackage{lipsum} % Used for inserting dummy 'Lorem ipsum' text into the template

\newcommand{\beq}{\begin{equation}}
\newcommand{\eeq}{\end{equation}}
\newcommand{\vv}{\mathbf{v}}

\newcommand{\bvec}{\begin{pmatrix}}
	\newcommand{\evec}{\end{pmatrix}}
\newcommand{\lp}{\left(}
\newcommand{\rp}{\right)}
\newcommand{\pa}[2]{\frac{\partial #1}{\partial #2}}
\newcommand{\paf}[2]{\partial #1 / \partial #2}

\newcommand{\ve}[1]{\mathbf{#1}}

\newcommand{\Wt}{\mathcal{W}}

\begin{document}

%----------------------------------------------------------------------------------------
%	TITLE PAGE
%----------------------------------------------------------------------------------------

%\begin{titlepage}

\title{Magnetogenesis by Wave-Driven Momentum Exchange}

\author{Ian E. Ochs}
\email{iochs@princeton.edu}
\affiliation{Department of Astrophysical Sciences, Princeton University, Princeton, New Jersey 08540, USA}
\author{Nathaniel J. Fisch}
\affiliation{Department of Astrophysical Sciences, Princeton University, Princeton, New Jersey 08540, USA}

%\author{Nathaniel J. Fisch}
%\affiliation{Department of Astrophysical Sciences, Princeton University, Princeton, New Jersey 08540, USA}
%\affiliation{Princeton Plasma Physics Laboratory, Princeton, New Jersey 08540, USA}


\begin{abstract}
When multiple species interact with an electrostatic ion acoustic wave, they can exchange momentum, despite the lack of momentum in the field itself.
The resulting force on the electrons can have a curl, and thus give rise to compensating electric fields with curl on magnetohydrodynamic timescales.
As a result, a magnetic field can be generated.
Surprisingly, in some astrophysical settings, this mechanism can seed magnetic fields with growth rates even larger than through the traditional Biermann battery.

\end{abstract}

%\maketitle
%----------------------------------------------------------------------------------------
%	TABLE OF CONTENTS
%----------------------------------------------------------------------------------------

%\tableofcontents % Include a table of contents

%\newpage % Begins the essay on a new page instead of on the same page as the table of contents 

%----------------------------------------------------------------------------------------
%	INTRODUCTION
%----------------------------------------------------------------------------------------

%\clearpage\newpage

\section{Introduction} 
Explaining the magnetic field structures present on different astrophysical scales is very difficult.
The observed magnetic fields in the universe are thought to be largely the result of amplification of small fields by magnetic dynamo mechanisms \citep{schober2011dynamoThesis,brandenburg2012current,squire2015dynamo,St-Onge2020}.
However, the dynamo requires a small, pre-existing ``seed'' magnetic field to act upon.
The generation of these seed fields on different scales is an area of active research.

Various mechanisms have been proposed as origins of these seed fields, from the Weibel instability \citep{schlickeiser2003cosmological}, to currents from charged cosmic rays \citep{miniati2011resistiveMagnetogenesis,Ohira2020}, to expulsion by jets from magnetized compact objects \citep{dalyLoeb1990jetMagnetogenesis}, to photon pressure on charged particles \citep{munirov2019CMB}.
However, the dominant favored mechanism in most scenarios  \citep{zweibel2013seeds,kulsrud1997magnetogenesis,gnedin2000magnetogenesisBiermann,hanayama2005biermann,hanayama2006snr,naoz2013biermannMagnetogenesis} is the Biermann battery \citep{biermann1950battery}.
The basic insight behind the mechanism is that electrons, with their negligible mass relative to ions, are inertia-free on magnetohydrodynamic (MHD) timescales.
Hence, the electrons must always be in force balance, and so a DC electric field must arise which cancels all other forces on the electrons.
If the applied non-DC force-per-electron $\ve{F}_e / n_e$ has curl, the induced electric field will also have curl, and give rise to a magnetic field via Faraday's law.

For the Biermann battery, the relevant force density is the pressure gradient, and a field is produced when the temperature and density gradients are misaligned.
Such nonaligned temperature and density gradients can be produced by shock-induced turbulence in the plasma, which can occur for instance when the expanding bubble of a supernova impacts an inhomogeneous interstellar medium (ISM) \citep{hanayama2005biermann, hanayama2006snr}.

Although the Biermann battery and literature that invokes it focuses on large-scale pressure forces, this is not the only force which can lead to Biermann-like magnetic induction.
In particular, we examine the forces resulting from wave-particle interactions.
Theory \citep{sagdeev1963shock,chen2012introduction} and experiments \citep{taylor1970IAWshock} show that shocks in an unmagnetized plasma form structures with trailing Debye-scale ion-acoustic waves (IAWs).
Thus, the same astrophysical shocked systems \citep{hanayama2005biermann,eichler1979shock} which give rise to Biermann generation are likely to give rise to IAWs and their associated forces.
Although the IAW, as an oscillating pressure force, does not lead to averaged Biermann field generation, as an electrostatic wave it can mediate directed momentum exchange between the electrons and ions along the direction of the wavevector \citep{Ochs2020Momentum}.
Thus, the IAW provides a net force on the electrons which, like the pressure gradient force in the Biermann mechanism, can have curl, and thus produce a magnetic field on MHD timescales.
As we show here, in some circumstances this mechanism, which we term the ``IAW battery,'' could lead to faster field growth than the Biermann battery.

\section{Biermann Battery from a Force with Curl}

The Biermann battery effect can be derived from Maxwell's equations and the electron momentum equation:
\begin{align}
	\pa{\ve{B}}{t} &= -c \nabla \times \ve{E} \label{eq:biermannFaraday}\\
	m_e n_e \frac{d\vv_e}{dt} &= -e n_e \lp \ve{E} + \frac{\vv_e}{c} \times \ve{B} \rp + \ve{F}_e. \label{eq:biermannElectronMomentum1}
\end{align}
Here, we use Gaussian units, $\ve{F}_e$ represents the force density due to all other forces on the electrons, and the remaining notation is standard.
To take the MHD limit, we consider a timescale long enough for the force on the electrons to equilibrate, which is equivalent to taking $m_e \rightarrow 0$. 
This makes Eq.~(\ref{eq:biermannElectronMomentum1}) an algebraic rather than differential equation, which we solve for $\ve{E}$:
\begin{align}
	\ve{E} &= - \frac{\vv_e}{c} \times \ve{B} + \frac{\ve{F}_e}{e n_e}. \label{eq:biermannMhdOhms}
\end{align}
This the Ohm's Law for our MHD model.
Taking $\vv_e \approx \vv_i \equiv \vv$, and plugging Eq.~(\ref{eq:biermannMhdOhms}) into Eq.~(\ref{eq:biermannFaraday}) then yields:
\begin{align}
\pa{\ve{B}}{t} = \nabla \times (\vv \times \ve{B}) - \frac{c}{e} \nabla \times \lp \frac{\ve{F}_e}{n_e}  \rp. \label{eq:biermannInduction}
\end{align}

For the Biermann battery, the relevant force is the electron pressure gradient force, $\ve{F}_e = -\nabla (n_e T_e)$.
Plugging this in to Eq.~(\ref{eq:biermannInduction}) yields the Biermann battery:
\begin{align}
\pa{B}{t}|_\text{Bier} &= \frac{c}{e n_e} \nabla n_e \times \nabla T_e.
\end{align}

In a typical astrophysical scenario, a shock will nonadiabatically heat the inhomogeneous interstellar medium (ISM), leading to nonaligned density and pressure gradients \citep{hanayama2005biermann}.
The growth rate of the field can then be estimated as:
\begin{align}
\pa{B}{t}|_\text{Bier} &\approx \frac{c T_e }{e L^2} \sin \theta, \label{eq:biermannEstimate}
\end{align}
where $\theta$ is the typical angle between the density and temperature gradients, and the scale length $L$ associated with the Biermann battery is the inhomogeneity scale length in the ISM, approximately 1-10 parsecs.

\section{Ion-Acoustic Wave Battery}

Shocks do not only form nonaligned pressure and density gradients: they can also produce ion-acoustic waves in the plasma \citep{sagdeev1963shock,chen2012introduction,taylor1970IAWshock}.
Thus, it is important to examine the effect of such waves on the generation of magnetic fields.

The electric field associated with a purely electrostatic plane wave such as an IAW has no momentum.
Therefore if the wave interacts with only one species, it cannot apply a net force as it damps.
However, a wave interacting with multiple species \emph{can} provide a net force to each species individually, as long as the forces on all species sum to zero.

An IAW in a plasma flattens the velocity distribution function in the neighborhood of the sound speed $C_s \equiv \sqrt{(Z T_e + T_i)/m_i}$, where $Z$ and $m_i$ are the ion charge state and mass respectively.
Because there tend to be more particles at low energy, the net effect is to accelerate particles to higher energy and momentum along the phase velocity.
This energy transfer energy from the wave into the particles is known as Landau damping.
To conserve momentum, the damping wave shifts the nonresonant velocity distribution in the opposite direction.
For an IAW, most of this nonresonant momentum transfer goes into the ions, so that both the electron and ion distributions experience a net force due to the wave.

The momentum transfer rate to the electrons for a narrow ion acoustic wave spectrum with $Z T_e \gg T_i$ is given by \cite{Ochs2020Momentum}:
\begin{align}
	\ve{F}_e &\approx \sqrt{\frac{\pi}{2}} \sqrt{\frac{Z m_e}{m_i}} \Wt \ve{k}, \label{eq:iawElectronForce}
\end{align}
where $\Wt$ is the energy in the ion acoustic wave (including the oscillating kinetic energy of the particles), $\ve{k}$ is the wave vector.

%This force cannot be applied forever: as the wave damps, the velocity distribution function flattens, and once it has flattened completely the momentum transfer stops.
%Thus, the maximum integrated force corresponds to completely flattening the distribution function in some vicinity $|v - C_s| < \epsilon C_s$ of the resonance, which is given by [cite paper 1]:
%\begin{align}
%	\lp \int \ve{F}_e dt\rp_{\max} &= \frac{\epsilon^3}{12\sqrt{2\pi}} \lp \frac{C_s}{v_{the}} \rp^3 (m_e n_e C_s),
%\end{align}
%where $v_{the} \equiv \sqrt{T_e / m_e}$ is the electron thermal velocity.

\begin{figure}[b]
	\center
	\includegraphics[width=0.8\linewidth]{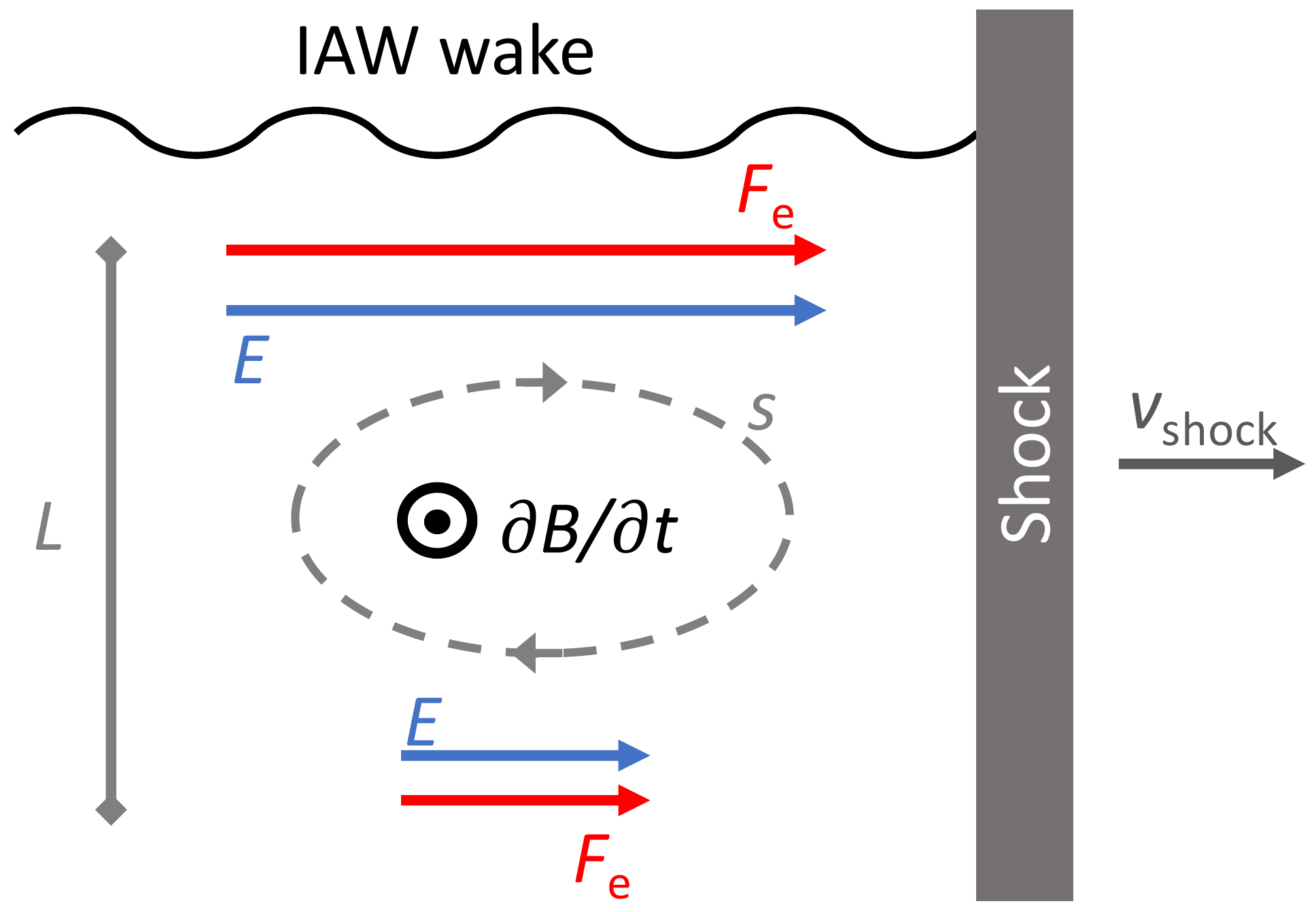}
	\caption{Mechanism of magnetogenesis by ion-acoustic waves.
		A shock propagates through the ISM, producing ion acoustic waves in its wake.
		These waves produce a force $\ve{F}_{e}$ electrons,  inhomogeneous on a scale $L$ in the ISM.
		A compensating inhomogeneous electric field $\ve{E}$ arises to cancel this force.
		This field has curl (consider the integral of the field over the loop $s$), and thus induces a magnetic field $\ve{B}$.
	}
	\label{fig:magnetogenesisMechanism}
\end{figure}

Inhomogeneities in a shocked plasma will naturally lead inhomogeneities in the wave spectrum generated by the shock, and so can produce an $\ve{F}_e$ with curl.
As in the Biermann battery, the resulting electron force will be compensated by an electric field with curl, and thus induce a magnetic field (Fig. \ref{fig:magnetogenesisMechanism}).
According to Eq.~(\ref{eq:biermannInduction}), and using the result from geometric optics for a narrow spectrum that $\nabla \times \ve{k} = 0$ \citep{dodin2012axiomatic}, the field growth rate will be:
\begin{align}
\pa{\ve{B}}{t}|_\text{IAW} = - \frac{c}{e} \sqrt{\frac{\pi}{2} \frac{Z m_e}{m_i}} \nabla \lp \frac{\Wt }{n_e} \rp \times \ve{k} . \label{eq:iawMagnetogenesis}
\end{align}
Thus, the scaling of the IAW battery is:
\begin{align}
	\pa{B}{t}|_\text{IAW} = \frac{c}{e} \sqrt{\frac{\pi}{2} \frac{Z m_e}{m_i}} \frac{\Wt k}{n_e L}. \label{eq:iawBatteryEstimate}
\end{align}

%This requires that knowing a characteristic correlation scale $L_c$ of the shock-produced IAW.
%From Ampere's law, this gives rise to a field $B_{QL} \sim j L_c/c$.
%Now, using Eq.~(\ref{eq:iawdjdtFinal}) and Eq.~(\ref{eq:iawWii}), we find:
%\begin{align}
%	\pa{B}{t}|_\text{QL} &\sim \Wt \frac{k L_c}{c}    \sqrt{\frac{Z m_e}{m_i}} \frac{e}{m_e}.
%\end{align}

We can get the ratio of the Biermann growth rate to the IAW growth rate simply by dividing Eq.~(\ref{eq:iawBatteryEstimate}) by Eq.~(\ref{eq:biermannEstimate}). 
Recalling that $k = 2\pi / \lambda_\text{IAW}$, where $\lambda_\text{IAW}$ is the typical wavelength of an ion acoustic wave, we arrive at our estimate of the relative strengths:
\begin{align}
	\frac{\paf{B}{t}|_\text{IAW}}{\paf{B}{t}|_\text{Bier}} \sim  \lp \frac{\sqrt{2 \pi^3}}{\sin \theta}\sqrt{\frac{ Z m_e}{m_i}} \rp \lp \frac{\Wt}{P_e} \rp \lp \frac{L}{\lambda_\text{IAW}} \rp.
\end{align}
Here, the first factor is $\mathcal{O}(10^{-1})$ at $\sin \theta = 1$, and can be significantly larger if the Biermann-relevant temperature and density gradients are closely aligned.
The second factor is $\mathcal{O}(1)$, if the wave energy is in equipartition with the thermal energy.
The final factor is the number of IAW wavelengths in a correlation scale.
There seems to be a great deal of uncertainty around the wavelength of shock-trailing IAWs in astrophysical settings, which could range from the experimentally-consistent electron Debye length, i.e. 160 meters for a 100 eV plasma at 0.2 cm$^{-3}$ \citep{hanayama2005biermann,mckeeOstriker1977ism}, to the scale of several parsecs \citep{spitzer1982acoustic}.
Thus, the last term could be extremely large,
and so it is quite plausible for the IAW growth rate to dominate in some scenarios.

\section{Connection to current drive in laboratory plasmas}

%The ion-acoustic wave battery has been proposed as a magnetogenesis mechanism for astrophysical applications because it produces a force on electrons with curl serving as  a source for the magnetic field as described in Eq.~(4).  
%Importantly, this wave is excited by the astrophysical shocks that also produce non-aligned pressure and density gradients, and so may compete with the magnetogenesis mechanisms associated with those gradients.  
An advantage of the IAW magnetogenesis mechanism is that the wave field itself need not carry any momentum, since it can drive current by catalyzing the exchange of momentum between electrons and ions.   
It is worthwhile to note that magnetogenesis by waves that themselves carry no momentum has been recognized both in theory and laboratory plasma experiments \citep{fisch1987theory}.  
For example, the electron cyclotron wave induces asymmetric collisions between electrons and ions \citep{fisch1980creating} to drive current.  

In the case of laboratory settings, magnetogenesis by waves has generally been termed  ``current drive,'' or ``radio-frequency (RF) current drive.'' 
This nomenclature arises, perhaps, because of the emphasis on maintaining steady state currents and their associated steady state magnetic fields in laboratory devices, rather than on the ab initio generation of the magnetic field.  
This steady state is maintained by the constant injection of RF wave power, which is balanced by collisional (resistive) dissipation of the current.  
However, the very same waves that maintain the steady state can, of course, also be used to generate the magnetic field.  
Thus, lower hybrid waves can maintain steady state currents \citep{Fisch1978}, but can also quite spectacularly generate large magnetic fields as well \citep{Fisch1985}.  

These RF current drive mechanisms, in an initial-value problem for the generation of the field, would enter in Eq.~(\ref{eq:biermannInduction}) through a force on electrons that is not curl free, much in the same way as the Biermann battery term or the ion acoustic wave battery term enters.
This approach accounts for the self-induction of the plasma which opposes the creation of the field, but which does not play a role in the eventual steady-state.
For the case of RF current drive, the field reaches a saturated steady state when the force term is balanced by collisional or resistive terms that do not appear in the collisionless limit of Ohm’s law as presented in Eq.~(\ref{eq:biermannMhdOhms}). 
If resistivity is  neglected, then other physical effects must be included to describe the saturation of the magnetogenesis, as described in the next section.
%Hence, if the counteracting effects of collisions are excluded, then these effects would have to be in the category of those effects that diminish the driving term, as described in the next section in discussing the saturation of the Biermann and  ion acoustic batteries.   

%\clearpage \newpage

\section{Saturation} 
A large growth rate is not sufficient to establish the IAW battery as a seeding mechanism for astrophysical magnetic fields.
As the battery proceeds, the plasma structure changes, and at some point the mechanism will saturate and the field production will cease. 
For the mechanism to be viable, the saturation level of the fields must be high enough to seed the astrophysical dynamo mechanisms---on the order of $10^{-20}-10^{-16}$ G for galactic magnetic fields \citep{Widrow2002}.

The first possible method of saturation, for either the Biermann battery or IAW battery, is the complete relaxation of the driving force.
For the Biermann battery, the pressure gradient should relax on approximately the sound crossing time $L / C_s$. 
Thus, integrating Eq.~(\ref{eq:biermannEstimate}) over this time, and taking $ZT_e \gg T_i$, we find:
\begin{align}
	B_\text{max,Bier} &\approx \frac{c m_i C_s}{e L} \sin \theta, \label{eq:biermannSaturation}.
\end{align}
This condition can be expressed more cleanly in terms of the ion cyclotron frequency $\Omega_{i,\text{Bier}} \equiv e B_\text{max,Bier} / c m_e$ associated with the saturated field, and the sound crossing time $L / C_s$ across the ISM scale length:
\begin{align}
	\Omega_{i,\text{Bier}}\lp \frac{L}{C_s} \rp &\lesssim \sin \theta. \label{eq:BiermannSaturation}
\end{align}
This expression assumes that the pressure force is thermal, rather than ram pressure; otherwise, the RHS will get an extra factor of $P_\text{ram} / n_e T_e$.

For the IAW battery, the driving force stops when the wave completely damps.
Integrating Eq.~(\ref{eq:iawBatteryEstimate}) over time, and using the results for a general wave that $\paf{\Wt}{t} = 2 \omega_i \Wt$, where for an electron-damped IAW $\omega_i = \omega \sqrt{\pi/8}  \sqrt{Z m_e / m_i}$, we find a similar result to that for the Biermann battery:
%\begin{align}
%B_\text{max,iaw} &\approx \frac{c m_i C_s}{e L} \frac{\Wt}{P_e}, \label{eq:iawSaturationAmp}.
%\end{align}
\begin{align}
\Omega_{i,\text{IAW}}\lp \frac{L}{C_s} \rp &\lesssim \frac{\Wt}{P_e}.
\end{align}
Thus, the ratio of the magnetic field saturation level in the Biermann vs. IAW battery is equal to the ratio of wave to thermal energy in the plasma.

However, for the IAW battery, there is a second saturation mechanism due to feedback from the magnetic field.
As the field grows in a plane perpendicular to $\ve{k}$, it will begin to influence the wave, preventing electron motion along $\ve{k}$.
The wave-particle interaction will be significantly impacted when the electron cyclotron frequency $\Omega_e$ becomes comparable to the wave frequency $\omega = C_s k$.
Thus, in addition to the constraint on the ion cyclotron frequency, we have a constraint on the electron cyclotron frequency:
\begin{align}
	\Omega_{e,\text{IAW}}\lp \frac{\lambda_\text{IAW}}{2 \pi C_s} \rp &\lesssim 1.
\end{align}

These constraints can be combined as:
\begin{align}
	\Omega_{i,\text{IAW}} \lp \frac{L}{C_s} \rp\lesssim %\begin{cases}
		 \min \lp \frac{\Wt}{P_e}, \frac{Z m_e}{m_i} \frac{\lambda_\text{IAW}}{2\pi} \rp.
	%\end{cases}
\end{align}
Thus, comparing to Eq.~(\ref{eq:BiermannSaturation}), we see that there should be many IAW wavelengths within the characteristic gradient scale length for the IAW battery to saturate at a similar level as the Biermann battery.

Finally, we can express the saturation field in number-form, which becomes:
\begin{align}
	B_{\text{IAW}} \lesssim \begin{cases}
		3.31 \times 10^{-17} \sqrt{\frac{\mu}{Z}} \frac{\Wt}{P_e} \lp \frac{T_e}{\text{1 eV}} \rp^{1/2} \lp \frac{L}{\text{1 pc}} \rp^{-1} \; \text{G}\\
		1.13 \times 10^{-19} \sqrt{\frac{Z}{\mu}} \lp \frac{T_e}{\text{1 eV}} \rp^{1/2} \lp \frac{\lambda_{\text{IAW}}}{\text{1 pc}} \rp^{-1} \; \text{G},
	\end{cases} 
\end{align}
where $\mu$ is the ion mass in a.m.u.
Thus, short-wavelength (relative to the ISM inhomogeneity scale length of $\sim 1$ pc) ion-acoustic waves in a shock-heated ($\sim$ 100 eV) ISM  could be able to seed galactic fields even at the level of $10^{-16}$ G.
As the wavelength becomes shorter and the hydrodynamic energy larger, the battery grows even stronger.

\section{Discussion}

There are some subtleties and caveats associated with the IAW battery.
In this section, we discuss in more detail the assumptions that go into the model, as well as their applicability.

First, an acoustic wave is formed by a set of oscillating pressure gradients, and yet we have demanded that the pressure gradient scale length $L$ be much greater than the acoustic wavelength $\lambda_{\text{IAW}}$.
This is consistent however; because the IAW field is oscillating, the resulting pressure force (and thus Biermann generation) will oscillate also, tending to cancel or at most grow as a random walk, $B \propto \sqrt{t}$. 
The corresponding IAW battery field, however, grows linearly with time.
Thus, the ion acoustic wavelength is a relevent scale length for the IAW field, but not the Biermann field.

%There are several requirements on the plasma in order for the IAW mechanism to work.
Second, the IAW battery requires that IAWs be only weakly damped by ions, i.e. $Z T_e \gg T_i$.
Thus, there must be either a source of electron heating, or some high charge states present in the plasma.

Third, the electron force term in Eq.~(\ref{eq:iawElectronForce}) applies to a Maxwellian plasma.
However, as the force is applied, the velocity distribution function will flatten in the neighborhood of the resonance, weakening the force.
Collisions between electrons must balance this flattening near resonance for a force to continue to be applied.
However, collisions between electrons and ions will add resistivity to the plasma, relaxing the generated field.
Thus, there must be enough collisions to keep the distribution function approximately Maxwellian near resonance, but not so many that the field diffuses out.
This is likely to be the case in the shock-heated ISM, where the collision time is on the order of years to decades for $(n, T) \sim (10^{-2} \text{ cm}^{-3}, 10^2 \text{ eV})$ \citep{mckeeOstriker1977ism}, while the dynamical timescales for e.g. the expansion of a supernova remnant are on the order of $10^5$ years \citep{hanayama2005biermann}.

%Of course, other effects could limit the efficacy of the IAW magnetogenesis mechanism.
%Third, the specific details of the wave spectrum could be modified in a non-Maxwellian plasma, although the basic asymmetries, wavenumbers, and resulting fields will likely remain similar.

Finally, the growth rate of this mechanism must of course be compared to other wave-driven mechanisms for any specific scenario, such as the Weibel instability \citep{schlickeiser2003cosmological}.

%\begin{align}
%	\frac{\paf{B}{t}|_\text{QL}}{\paf{B}{t}|_\text{Bier}} \sim  k L_c \frac{\Wt}{P_e} \frac{\omega_{pe} \omega_{pi} L_B^2}{c^2}.
%\end{align}
%Here, the first term is the ratio of the correlation length to the wavelength in the shock trail, likely large compared to one, and the second term is the ratio of wave to thermal energy, which is likely $\lesssim 1$.
%The third term then largely determines the dominant effect.
%This third term is a comparison of an artificial hybrid frequency, $\omega_H \sim \sqrt{\omega_{pe} \omega_{pi}}$, to the time for light to traverse the Biermann scale lengths.
%In fact, this factor is much greater than one, since the electron plasma frequency is typically on the order of $10^4$ Hz even in the ISM, while the transit time of light across the Biermann scale lengths is on the order of years.
%Thus, it seems likely that the ion-acoustic wake of the shock could lead to much faster field growth, on much smaller correlation scale lengths, than the Biermann term.

%\textbf{I think there might be more to this comparison. For instance, the Biermann term seems like it should be oscillatory in the presence of IAW wake turbulence, and thus produce a ``random walk'' growth. Meanwhile, the quasilinear growth occurs in a consistent direction on a larger correlation scale.}

\section{Conclusion}
We showed how wave-driven momentum exchange could provide a magnetogenesis mechanism similar to the Biermann battery in astrophysical settings, and how this mechanism could potentially be stronger than the Biermann mechanism in certain scenarios.
Unusually, it is a kinetic mechanism that produces fields on  hydrodynamic length scales.
As a mechanism that is based on long-established, experimentally-verified plasma physics models, the IAW battery is an attractive candidate for magnetogenesis in astrophysical settings.

\acknowledgments
We would like to thank E.J. Kolmes and M.E. Mlodik for helpful discussions.
This work was supported by grants  DOE DE-SC0016072 and   DOE NNSA DE-NA0003871.
One author (IEO) also acknowledges the support of the DOE Computational Science Graduate Fellowship (DOE grant number DE-FG02-97ER25308).

%\clearpage

%\bibliographystyle{unsrt}

%\bibliography{../../../Reading/allRefsMend.bib}
%\bibliography{comprehensive,More}
\clearpage\newpage

\end{document}